\documentclass[conference]{IEEEtran}
\IEEEoverridecommandlockouts
\usepackage{multirow}
\usepackage{boldline,multirow}
\usepackage{cite}
\usepackage{amsmath,amssymb,amsfonts}
\usepackage{algorithm}
\usepackage{algpseudocode}
\usepackage{graphicx}
\usepackage{textcomp}
\usepackage{mathtools}
\usepackage{xcolor}
\usepackage{booktabs}
\usepackage{amsmath}
\usepackage{array} 
\usepackage{subcaption, multicol}
\usepackage{bm}
\usepackage{url}
\usepackage{hyperref}
\usepackage{diagbox}
\usepackage[english]{babel}
\usepackage{amsthm, nccmath}
\usepackage{stackengine}
\algnewcommand{\IIf}[1]{\State\algorithmicif\ #1\ \algorithmicthen}
\algnewcommand{\ElseIIf}[1]{\algorithmicelse\ #1} 
\algnewcommand{\EndIIf}{\unskip\ \algorithmicend\ \algorithmicif}

\DeclarePairedDelimiter\ceil{\lceil}{\rceil}

\stackMath
\newlength\matfield
\newlength\tmplength
\def\matscale{1.}
\newcommand\dimbox[3]{%
  \setlength\matfield{\matscale\baselineskip}%
  \setbox0=\hbox{\vphantom{X}\smash{#3}}%
  \setlength{\tmplength}{#1\matfield-\ht0-\dp0}%
  \fboxrule=1pt\fboxsep=-\fboxrule\relax%
  \fbox{\makebox[#2\matfield]{\addstackgap[.5\tmplength]{\box0}}}%
}

\newcommand\matbox[5]{
  \stackunder{\dimbox{#1}{#2}{$#5$}}{\scriptstyle(#3\times #4)}%
}
\theoremstyle{definition}

\newcolumntype{C}[1]{>{\centering\arraybackslash}p{#1}}
\def\BibTeX{{\rm B\kern-.05em{\sc i\kern-.025em b}\kern-.08em
    T\kern-.1667em\lower.7ex\hbox{E}\kern-.125emX}}
\begin{document}
\title{Theoretical Analysis of the Efficient-Memory Matrix Storage Method for Quantum Emulation Accelerators with Gate Fusion on FPGAs \\

}


\author{
	 \IEEEauthorblockN{Le Tran Xuan Hieu \textsuperscript{1,2}, Pham Hoai Luan \textsuperscript{3}, Vu Tuan Hai \textsuperscript{3}, Le Vu Trung Duong \textsuperscript{3} and Yasuhiko Nakashima\textsuperscript{3}}
	\IEEEauthorblockA{
    \textsuperscript{1} University of Information Technology, Ho Chi Minh City, 700000, Vietnam.\\ \textsuperscript{2} Vietnam National University, Ho Chi Minh City, 700000, Vietnam.\\ \textsuperscript{3} Nara Institute of Science and Technology, 8916–5 Takayama-cho, Ikoma, Nara 630-0192, Japan.\\ 
Email: pham.luan@naist.is.jp} 
}

\maketitle

\begin{abstract}

Quantum emulators play an important role in the development and testing of quantum algorithms, especially given the limitations of the current FTQC era. Developing high-speed, memory-optimized quantum emulators is a growing research trend, with gate fusion being a promising technique. However, existing gate fusion implementations often struggle to efficiently support large-scale quantum systems with a high number of qubits due to a lack of optimizations for the exponential growth in memory requirements. Therefore, this study proposes the EMMS (Efficient-Memory Matrix Storage) method for storing quantum operators and states, along with an EMMS-based Quantum Emulator Accelerator (QEA) architecture that incorporates multiple processing elements (PEs) to accelerate tensor product and matrix multiplication computations in quantum emulation with gate fusion. The theoretical analysis of the QEA on the Xilinx ZCU102 FPGA, using varying numbers of PEs and different depths of unitary and local data memory, reveals a linear increase in memory depth with the number of qubits. This scaling highlights the potential of the EMMS-based QEA to accommodate larger quantum circuits, providing insights into selecting appropriate memory sizes and FPGA devices. Furthermore, the estimated performance of the QEA with PE counts ranging from $2^2$ to $2^5$ on the Xilinx ZCU102 FPGA demonstrates that increasing the number of PEs significantly reduces the computation cycle count for circuits with fewer than 18 qubits, making it significantly faster than previous works.

\end{abstract}



\section{Introduction}
\label{sec:introduction}

Nowadays, the use of quantum algorithms on FTQC computers has great potential for speeding up computation \cite{lau2022nisq, Acharya2023, Bluvstein2024}. The expense of performing tests on real quantum computers via cloud services such as AWS Braket, however, is an obstacle. As a result, scientists frequently turn to employing simulation programs such as Qiskit \cite{javadiabhari2024quantumcomputingqiskit} and Pennylane \cite{bergholm2022pennylaneautomaticdifferentiationhybrid} to test their models and methods locally. Although this method can be used for preliminary verification, it is inefficient since it takes exponentially more computer power to simulate and store the state of a large number of qubits ($\#\text{Qubits}$). Moreover, the functions performed by a quantum simulator always entail massive matrix multiplications and tensor products. The dimensions of these matrices rise exponentially with the $\#\text{Qubits}$, posing significant challenges with processing time and memory usage. A potential solution to these challenges is the growing use of quantum emulators built on specialized hardware. These emulators take advantage of the hardware’s strengths, such as enhanced parallel processing and more flexible memory management, allowing them to efficiently simulate larger qubit systems. By focusing on specialized hardware, these emulators can drive long-term progress, enabling quicker discovery and testing of quantum algorithms.

Numerous studies have been conducted to enhance the performance of quantum circuit simulation systems, aiming to achieve high simulation speeds and support computations for a larger $\#\text{Qubits}$. For example, the study in \cite{H_ner_2017} used the Cori II supercomputer, with $0.5$ PBs of memory, to efficiently boost the simulation speed for a $45$-qubit quantum circuit simulator using scheduling algorithms and automated code generation. However, its huge memory demands limited scalability and accessibility for larger quantum circuits. Another emerging trend is the use of graphics processing units (GPUs) for quantum circuit modeling acceleration. As described in \cite{10313722}, the NVIDIA cuQuantum SDK provides useful building blocks for tensor network and state vector-based simulators to accelerate computations by leveraging the parallel processing power of GPUs. Another work presented in \cite{9773245} focuses on GPU-based simulation optimization using active state amplitude modulation, pruning, and compression techniques. Despite achieving high performance, they still can not optimize memory usage effectively, making it difficult to support quantum simulation systems with a large $\#\text{Qubits}$. Notably, FPGAs have emerged as a promising hardware platform for quantum circuit emulations. For instance, the efficient computation quantum emulator for unitary transformations proposed in \cite{mahmud2020efficient} utilizes a pipeline design, floating-point arithmetic, and hardware minimization techniques to simulate Quantum Fourier Transform (QFT) \cite{wakeham2024inferenceinterferenceinvariancequantum} with a high $\#\text{Qubits}$. This approach aims to improve processing speed, accuracy, and memory usage. In addition, the re-configurable quantum emulator proposed in \cite{Mahmud2020Grover} modifies Grover's algorithm for dynamic multi-pattern search and estimates the capability of simulating up to 32 qubits with high accuracy and resource optimization techniques. However, the scalability of these FPGA-based emulators is still limited by the large hardware resources required.  Furthermore, the simulation time increases exponentially with $\#\text{Qubits}$, and efficiently simulating complex algorithms can be challenging.

A potential solution to the problem of enormous storage requirements in quantum simulators is gate fusion, as proposed and optimized in \cite{H_ner_2017,weko_210570_1,smelyanskiy2016qhipster}. Accordingly, this technique reduces the number of gates in a quantum circuit by merging multiple gates into a single gate, which in turn reduces the number of matrix multiplications needed to simulate the circuit. As a result, memory consumption is decreased, and simulation performance is increased. Gate fusion can be very helpful in reducing the number of matrix operations, but it might not be able to completely address the memory use problem, particularly when working with high $\#\text{Qubits}$. The prior quantum states also take a significant amount of memory to store, and the aggregated matrices can still be huge.

To address the existing challenges, this paper introduces the Efficient-Memory Matrix Storage (EMMS) method, designed to minimize storage requirements and maximize parallel processing speed for gate fusion quantum simulation. Furthermore, a quantum emulation accelerator (QEA) based on the EMMS method is proposed to be implemented on FPGA system-on-chips (SoCs) to demonstrate the effectiveness of the method. The EMMS's effectiveness in reducing storage resources and improving processing speed is exhibited through evaluation results incorporating variations in memory resources, $\#\text{PEs}$ and the $\#\text{Qubits}$.

\section{Background knowledge} \label{sec:background}

\subsection{Quantum computation model}
\label{sec:quantum_computation}

\begin{figure}[t]
    \centering
    \includegraphics[width=0.49\textwidth]{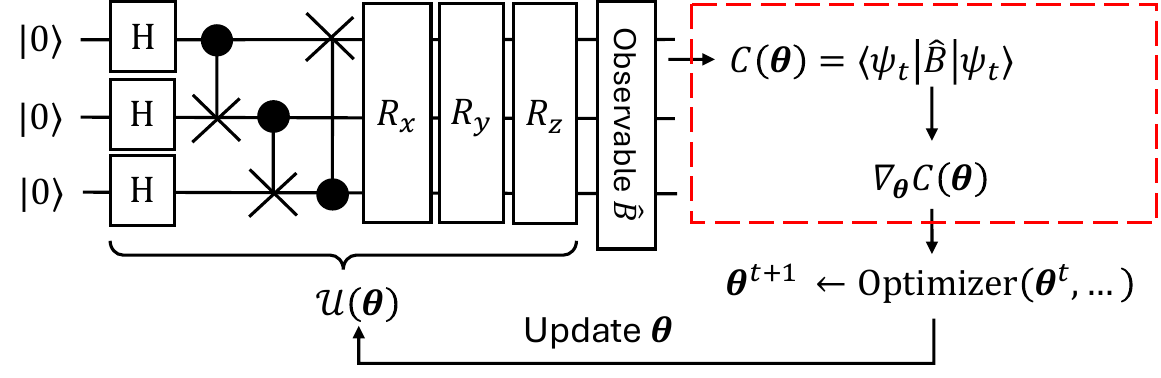}
    \caption{Computational path in quantum computation model.}
    \label{fig:pie}  
\end{figure}

The quantum computation model can be expressed as a unitary $\mathcal{U}$ parameterized by $\bm\theta$, act on a reference state $|\psi^{\bm{(0)}}\rangle$ \cite{schuld2021machine} as Eq.~\ref{eq:basic}:

\begin{equation}
    \begin{split}
        |\psi^{(\bm{m})}\rangle = \mathcal{U}(\bm\theta)\times|\psi^{(\bm{0})}\rangle
        = \left(\prod_{j=1}^m U^{(\bm{t})}(\theta^{(\bm{t})})\right)\times|\psi^{(\bm{0})}\rangle,
    \end{split}
    \label{eq:basic}
\end{equation}

where $|\psi^{(m)}\rangle$ represents the target quantum state, and $\mathcal{U}(\bm\theta)$ denotes the quantum operator. Whenever the quantum computation model is performed, it constitutes a Quantum Evaluation (\textbf{QE}). This quantum operator can be decomposed not only into Matrix Multiplication (\textbf{MM}) between sub-operators but also into $\hat{m}$ gates (considering single-qubit gates and two-qubit gates) application. These gates $G=\{g_j\}$ can be applied on $|\psi^{(\bm{0})}\rangle$ sequentially to achieve the same $|\psi^{\bm{(m)}}\rangle$.



The strategy for solving Eq.~\ref{eq:basic} is choosing the break points in gate sequence $g_1 g_2 \ldots g_{\hat{m}}$ to break it as group $\{G^{(\bm{t})}\}$, satisfy single parameter condition (each sub-operator $U^{(\bm{t})}(\theta^{(\bm{t})})$ has as many gates as possible but only contains one parameterized gate), then pass $\{G^{(\bm{t})}\}$ into Tensor Product (\textbf{TP}) function $\mathcal{T}: G^{(\bm{t})}\in g^{k} \rightarrow U^{(\bm{t})}\in \mathbb{C}^{N \times N} = \bigotimes_{j=1}^{|G^{(\bm{t})}|}g_j$ with $g_j\in G^{(\bm{t})}$. The term $U^{(\bm{m})} \ldots U^{(\bm{1})} |\psi^{(\bm{0})}\rangle$ can be re-written as Eq.~\ref{eq:basic2}:

\begin{equation}
    \begin{split}
        \left(\mathcal{T}(G^{(\bm{m})})\left(\ldots\left(\mathcal{T}(G^{(\bm{1}}))|\psi^{(\bm{0})}\rangle\right)\right)\right),
    \end{split}
    \label{eq:basic2}
\end{equation}

which require $m$ $\times$ \textbf{TP} and $m$ $\times$ \textbf{MM} for execution. Note that, $U^{(\bm{t})}$ is the unitary matrix, satisfy $U^{(\bm{t})}(U^{(\bm{t})})^{\dagger} = \mathbb{I}$.

\subsection{Sparse matrix computations}  \label{subsec:smc}

    \subsubsection{Coordinate Format}
        A matrix is typically regarded as sparse when the number of zero elements is greater than or equal to the number of non-zero elements. In the context of quantum emulation, the enormous size of these matrices poses a challenge, as they contain so many elements that conventional memory systems cannot accommodate them. To efficiently manage this, we employ a sparse matrix storage method known as the Coordinate (COO) Format \cite{MIT_COO}. This format focuses on storing only the non-zero values, thereby significantly reducing the amount of storage required. By adopting the COO Format, we also exclude the need for redundant computations involving zero values, which in turn leads to enhanced computational performance. The format achieves this efficiency by representing a sparse matrix using three primary components: row indices, column indices, and the actual non-zero values. Each non-zero element is mapped precisely to its location within the tensor using these indices, allowing for efficient data retrieval and manipulation.
        
    \subsubsection{Tensor product} \label{subsubsec:tenmul}

        One of the computations in quantum emulation is \textbf{TP} between two quantum operators, also known as the Kronecker product. Denoted by $g \otimes h$, it is an operation on two matrices of arbitrary size resulting in a block matrix.
        
\begin{figure}[t]
            \centering
            \includegraphics[width=0.49\textwidth]{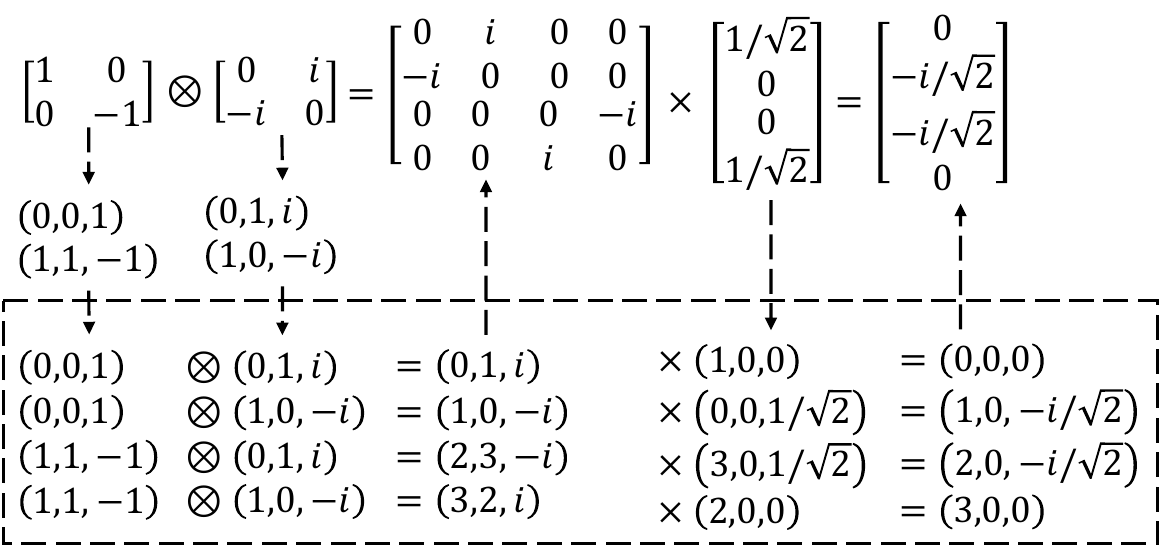}
            \caption{Conducting both \textbf{TP} and \textbf{MM} $(\mathcal{T}(\{Z, Y\})|\psi_{\text{GHZ}}\rangle)$ in COO format.}
            \label{fig:coomul}  
        \end{figure}

    
        As we discussed, COO format is not only used for storing sparse matrices but also for computing. Accordingly, each non-zero element $g_{i,j}\in g$ is denoted as $(i,j,g_{i,j})$ where $i,j, g_{i,j}$ are the row index, column index, and value, respectively. The \textbf{TP} is expressed as looping through all possible combinations between $\{g_{i,j}\}$ and $\{h_{k,l}\}$, as Algo.~\ref{algo:tensor}. Fig.~\ref{fig:coomul} illustrates the detailed sparse \textbf{TP} between two $2 \times 2$ tensors, which are transformed into two corresponding arrays of tuples. The resulting computed tuples are then merged into a new array.

        \begin{algorithm}
        \caption{\textbf{TP} between two gates in COO.} 
        \label{algo:tensor}
        \begin{algorithmic}[]
        \Require $g, h$, $\text{res}\gets \bm{0}^{(|g|\times |h|) \times (|g|\times |h|)}$
        \For{$(i,j,g_{i,j})$ in $g$}
            \For{$(k,l,h_{k,l})$ in $h$}
                \State $\text{res}_{i \times |h| + l, j \times |h| + k} \gets g_{i,j}\times h_{k,l}$
            \EndFor    
        \EndFor
        \State \Return $\text{res}$
        \end{algorithmic}
        \end{algorithm}

   \subsubsection{Matrix multiplication} \label{subusbsec:maxmul}
        \begin{algorithm}
        \caption{\textbf{MM} between operator and state in COO.} 
        \label{algo:matmul}
        \begin{algorithmic}[]
        \Require $U^{(\bm{t})}, |\psi^{(\bm{t-1})}\rangle$, $\text{res}\gets \bm{0}^{n\times 1}$
        \For{$(i,j,U^{(\bm{t})}_{i,j})$ in $U^{(\bm{t})}$}
            \For{$(k,0,\alpha_{k})$ in $|\psi^{(\bm{t - 1})}\rangle$}
                \IIf{$j == k$} $\text{res}_{i} \gets \text{res}_{i} + U^{(\bm{t})}_{i,j}\times \alpha_{k}$\EndIIf
            \EndFor    
        \EndFor
        \State \Return $\text{res}$
        \end{algorithmic}
        \end{algorithm}
        
        In addition to the \textbf{TP}, quantum emulation employs the \textbf{MM} working as presented in Algo.~\ref{algo:matmul}. Similarly, the COO format is utilized for this computation, where all non-zero elements of matrices will be represented as constant or non-constant tuples as described in \ref{subsubsec:tenmul}. When performing $U^{(\bm{t})} \times |\psi^{(\bm{t-1})}\rangle$, non-zero elements of $U^{(\bm{t})}$ are multiplied with corresponding non-zero elements of $|\psi^{(\bm{t-1})}\rangle$ based on ESC algorithm \cite{dalton2015optimizing}.
        Specifically, each tuple $(i,j,U^{(\bm{t})}_{i,j})$ is computed with each amplitude $\alpha_k$, provided that $j$ = $k$. All tuples with the same $(i, k)$ are accumulated to obtain the final result as shown in Fig.~\ref{fig:coomul}.

        
        
        
        

\section{Proposed Method and Hardware Architecture}
This section introduces the Efficient-Memory Matrix Storage method to reduce memory usage and improve processing speed, along with our proposed architecture for efficient quantum emulation on FPGAs.

\subsection{Efficient-Memory Matrix Storage (EMMS) Method} \label{subsec:EMMS}

\begin{figure}[t]
    \centering
    \includegraphics[width=0.49\textwidth]{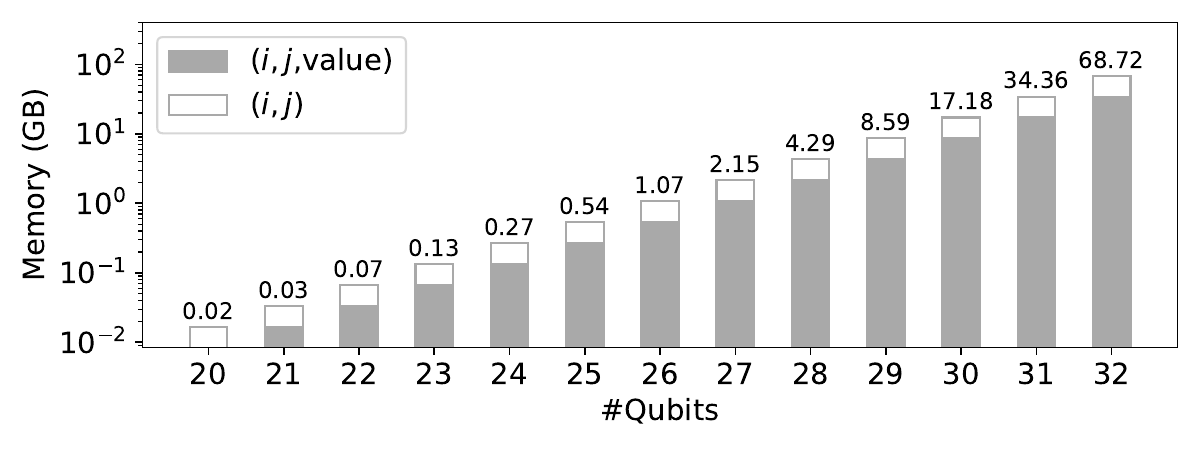}
    \caption{Traditional gate fusion ($U^{(\bm{t})}$) storage method in \cite{weko_210570_1}.}
    \label{fig:graph_traditional}
\end{figure}

The challenge when simulating Eq.~\ref{eq:basic} is the exponential scaling of both computation time and resources based on ($\#\text{Qubits}$) $n$. In detail, $\{U_j\}$ can be represented as $N\times N$ complex matrix ($N = 2^n)$. Fig.~\ref{fig:graph_traditional} illustrates the relationship between the $\#\text{Qubits}$ in a quantum circuit and the memory usage (in GB) for $U^{(\bm{t})}$. It is observed that the memory requirements for simulating quantum circuits scale exponentially with the $\#\text{Qubits}$. This phenomenon becomes more pronounced as the $\#\text{Qubits}$ increases. For instance, simulating a circuit with 20 qubits requires 0.02 GB of memory. However, as the qubit count increases to 32, the memory requirements skyrocket to approximately 68.72 GB. In almost all simulators, only $|\psi^{(\bm{t})}\rangle$ is being saved during the whole computation process. Our target is to reducing both computation time and resources to polynomial complexity at a certain $\#\text{Qubits}$.

The challenge for processing $U^{(\bm{t+1})} = \mathcal{T}(G^{(\bm{t+1})})$ is that the output size is a $N \times N$ matrix, which is soon to be overloaded for local memory. Our strategy is decomposing $\mathcal{T}(G^{(\bm{t+1})})$ into $\mathcal{T}(\overline{G}) \otimes \mathcal{T}(G)$ as Eq.~\ref{eq:explicit}. With $\overline{n}$ ($\overline{N} = 2^ {\overline{n}}$) is the dividing point which depends on the type of quantum gates and the hardware resources, detailed in Section~\ref{subsec:architecture}. The gate-chosen strategy is priority grouping sparse gates (all gates excluding $\{H,R_X(.), R_Y(.),SX\}$), for dense gates, we group it with $\mathbb{I}$. This leads to the number of non-zero elements in each row in $\mathcal{T}(\overline{G})$ is $1$.


\begin{equation}
\begin{split}
    \mathcal{T}(G^{(\bm{t+1})})|\psi^{(\bm{t})}\rangle&= \left[ \mathcal{T}(\overline{G}) \otimes \mathcal{T}(G)\right] |\psi^{(\bm{t})}\rangle \\
    &= \left[\left(\bigotimes_{j=0}^{\overline{n} - 1}g_j\right) \otimes \left(\bigotimes_{j=\overline{n}}^{n - 1}g_j\right) \right] \begin{bmatrix}
           \alpha^{(\bm{t})}_{0} \\
           \alpha^{(\bm{t})}_{1} \\
           \vdots \\
           \alpha^{(\bm{t})}_{N}
         \end{bmatrix}
\end{split}
\label{eq:explicit}
\end{equation}

\begin{equation*}
\resizebox{0.49\textwidth}{!}{
$ 
\begin{split}
    &=\left[
\matbox{7}{5}{\overline{N}}{1}{\{(i,j,\overline{G}_{i,j})\}} \otimes \matbox{7}{9}{\frac{N}{ \overline{N}}}{1}{
    \begin{array}{c} (0,j_0,G_{0,j}) \\ (1,j_1,G_{1,j}) \\ \vdots \\(\frac{N}{\overline{N}}-1,j_{\frac{N}{\overline{N}} - 1},G_{\frac{N}{\overline{N}}-1,j})\end{array}} \right]
    \\ &\times                  \begin{bmatrix}
            \alpha^{(\bm{t})}_0 \\
           \vdots \\
           \underbrace{\begin{bmatrix}
           \alpha^{(\bm{t})}_{j \frac{N}{ \overline{N}}} \\
           \alpha^{(\bm{t})}_{j \frac{N}{ \overline{N}} + 1} \\
           \vdots \\
           \alpha^{(\bm{t})}_{(j + 1)\frac{N}{ \overline{N}} - 1}
           \end{bmatrix}}_{|\psi^{(\bm{t})}\rangle_{j \frac{N}{ \overline{N}}:(j+1)\frac{N}{ \overline{N}}-1}} \\
           \vdots \\
           \alpha^{(\bm{t})}_{N-1} \\
         \end{bmatrix}
    =      \begin{bmatrix}
           \alpha^{(\bm{t + 1})}_0 \\
           \vdots \\
           \underbrace{\begin{bmatrix}
           \alpha^{(\bm{t + 1})}_{i \frac{N}{ \overline{N}}} \\
           \alpha^{(\bm{t + 1})}_{i \frac{N}{ \overline{N}} + 1} \\
           \vdots \\
           \alpha^{(\bm{t + 1})}_{(i + 1)\frac{N}{ \overline{N}} - 1} 
           \end{bmatrix}}_{|\psi_{\bm{t + 1}}\rangle_{i \frac{N}{ \overline{N}}:(i+1)\frac{N}{ \overline{N}}-1}} \\
           \vdots \\
           \alpha^{(\bm{t + 1})}_{N-1}
         \end{bmatrix}
\end{split}
$}
\end{equation*}


The right side of Eq.~\ref{eq:explicit} ($\left[ \mathcal{T}(\overline{G}) \otimes \mathcal{T}(G)\right] |\psi^{(\bm{t})}\rangle$) is executed by reiterating \textbf{TP} and \textbf{MM} through all tuples in $\mathcal{T}(\overline{G})$ as Algo.~\ref{algo:basic}, instead of Algo.~\ref{algo:matmul}. Note that multiple PEs can handle this process parallelly. A more comprehensive example is presented in Appendix.~\ref{sec:example}.

\begin{figure}[t]
    \centering
    \includegraphics[width=0.49\textwidth]{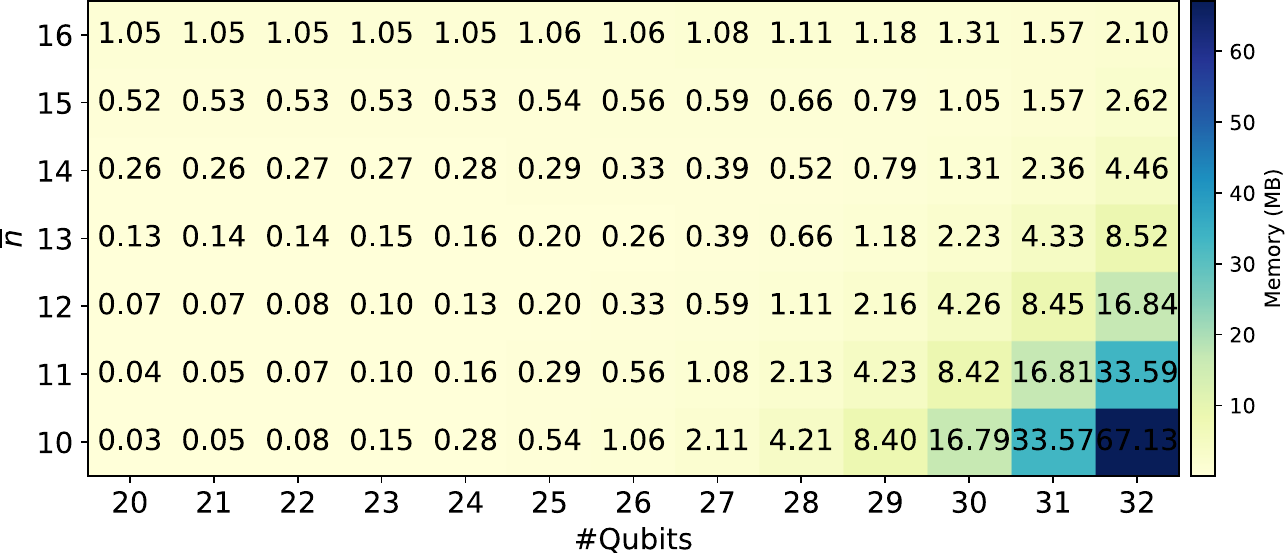}
    \caption{Proposed gate fusion's storage.}
    \label{fig:graph_MMS}  
\end{figure}

This methodology effectively reduces memory usage by removing redundant zero elements using the COO format. Furthermore, decomposing $\mathcal{T}(G^{(\bm{t+1})})$ into $\mathcal{T}(\overline{G})$ and $\mathcal{T}(G)$ reduces the storage requirements from $N$ to $\overline{N} + \frac{N}{\overline{N}}$. Under ideal conditions, where $\overline{n} = \ceil{n/2}$, the expression $\overline{N} + \frac{N}{\overline{N}}$ simplifies to $2\overline{N}\ll \overline{N}^{2}$, resulting in a significant memory efficiency improvement of $2^{\overline{n} - 1}$. Fig.~\ref{fig:graph_MMS} shows the estimated memory space from $20$ to $32$ qubits with different $\overline{n}$. The optimal cases lie on the diagonal line. It is observed that selecting an appropriate value for $\overline{n}$ allow the EMMS method to limit the storage usage to a maximum of 3 MB.

\begin{algorithm}
\caption{Basic state-evolution operation} 
\label{algo:basic}
\begin{algorithmic}[]
\Require $|\psi^{(\bm{t})}\rangle,|\psi^{(\bm{t+1})}\rangle\gets [0,0,\ldots, 0]$

\State 
\For{$i$ in $[0,1,\ldots,\overline{N}-1]$}
    \State $|\psi^{(\bm{t+1})}\rangle_{i\frac{N}{\overline{N}}:(i+1)\frac{N}{\overline{N}}-1} \gets \left[\overline{G}_{i, j}\times \mathcal{T}(G)\right] \times |\psi^{(\bm{t})}\rangle_{j\frac{N}{\overline{N}}:(j+1)\frac{N}{\overline{N}}-1} $
\EndFor
\State \Return $|\psi^{(\bm{t+1})}\rangle$
\end{algorithmic}
\end{algorithm}

    
\subsection{EMMS based-Quantum Emulation Accelerators} \label{subsec:architecture}

According to the methodology discussed in Section~\ref{subsec:EMMS}, we present an overview of the hardware architecture, focusing on the memory organization for $\mathcal{T}(\overline{G})$, $\mathcal{T}(G)$, $|\psi^{\bm{t}}\rangle$ and $|\psi^{\bm{t+1}}\rangle$ as detailed in Section~\ref{subsubsec:architecture}. Additionally, the parallel execution of \textbf{TP} and \textbf{MM} is outlined in Section~\ref{subsubsec:PE}, while the design of a Complex ALU optimized for Sparse Matrix Operations is explained in Section~\ref{subsubsec:ALU}.

\subsubsection{Overview Hardware Architecture} \label{subsubsec:architecture}

    \begin{figure}[t]
        \centering
        \includegraphics[width=0.49\textwidth]{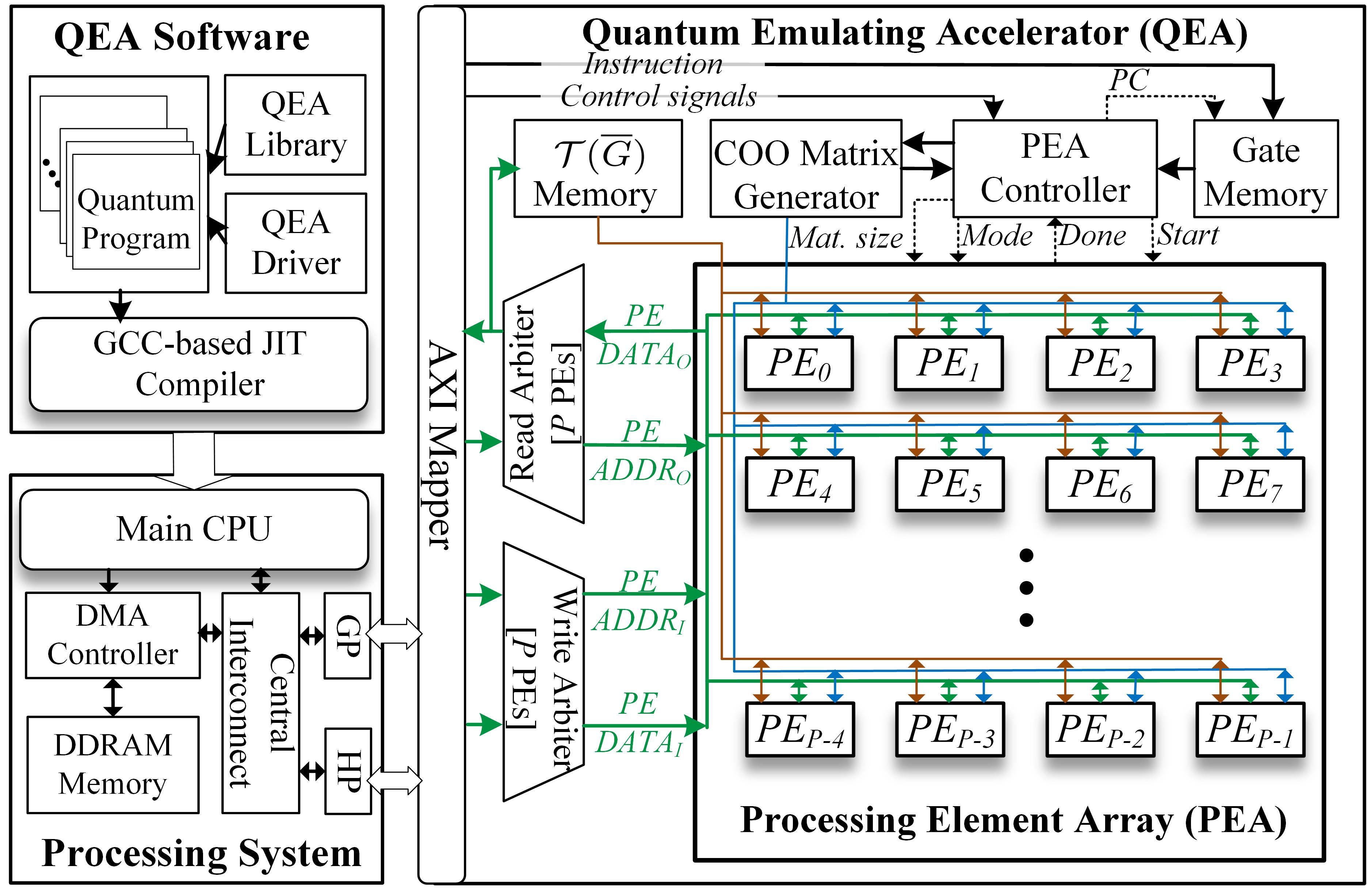}
        \caption{Hardware architecture of the quantum emulating accelerator.}
        \label{fig:overview}  
    \end{figure}

    Fig.~\ref{fig:overview} shows the overview architecture of a QEA at the system-on-chip (SoC) level, which consists of two main parts: the processing system (PS) and programmable logic (PL). The PS communicates with PL through a 128-bit AXI bus.
    
    The QEA's PS, separated into user and kernel space, is in charge of running the GNU/Linux operating system. Programmers can compute $\cos$ and $\sin$ values, as well as create a Instructions list of gate names and positions, using the user space. Configuration, control, and computational data of Instructions programs are compiled into machine code in the kernel space by an QEA compiler with GCC assistance. Through the use of direct memory access (DMA) or programmable input/output (PIO) over the AXI bus, the FPGA driver transfers the compiled data from the PS to the programmable logic (PL). Since control data, including start or finish signals are small and configuration data are transferred only once, it is sufficient to transfer both types of data using a 64-bit width PIO.
    
    The main component, QEA is realized within the PL of the FPGA and consists of seven parts: an AXI mapper, a $\mathcal{T}(\overline{G})$ Memory, a COO Matrix Generator, a PEA Controller, Gate Memory, a Write Arbiter, Read Arbiter and a Processing Element Array (PEA). The AXI Mapper oversees data exchange between the PS and the Local Data Memories of PEs. The Gate Memory is used to store the instructions received from the processing system (PS). The Processing Element Array (PEA) controller instructs the COO Matrix Generator to generate COO List gate matrices, as shown in Table~\ref{tab:c_gate}, using the instruction list stored in the Gate Memory. The matrix data is transferred to the PEA for the computation of $\mathcal{T}(\overline{G})$, which is then stored in the $\mathcal{T}(\overline{G})$ Memory. Simultaneously, this process also computes $\mathcal{T}(G),$ which is stored in the PEA. The value of $\mathcal{T}(\overline{G})$ is then transferred back to the PEA to compute $U$ and the $|\psi^{\bm{t}}\rangle$ stored in DDRAM is transferred to the PEA for calculating the next state, $|\psi^{(\bm{t+1})}\rangle$. This updated state is then returned to DDRAM for storage.


\subsubsection{Processing Element for Parallel \textbf{TP} and \textbf{MM} Computations} \label{subsubsec:PE}

    Despite the reduction in memory requirements, the computational demands of matrix operations remain substantial. To address this challenge, we integrate processing elements (PEs) with local data memory. This architecture ensures efficient execution of \textbf{MM} and \textbf{TP} operations, facilitating parallel processing and reducing overall latency.
    
    \begin{figure}[t]
        \centering
        \includegraphics[width=0.47\textwidth]{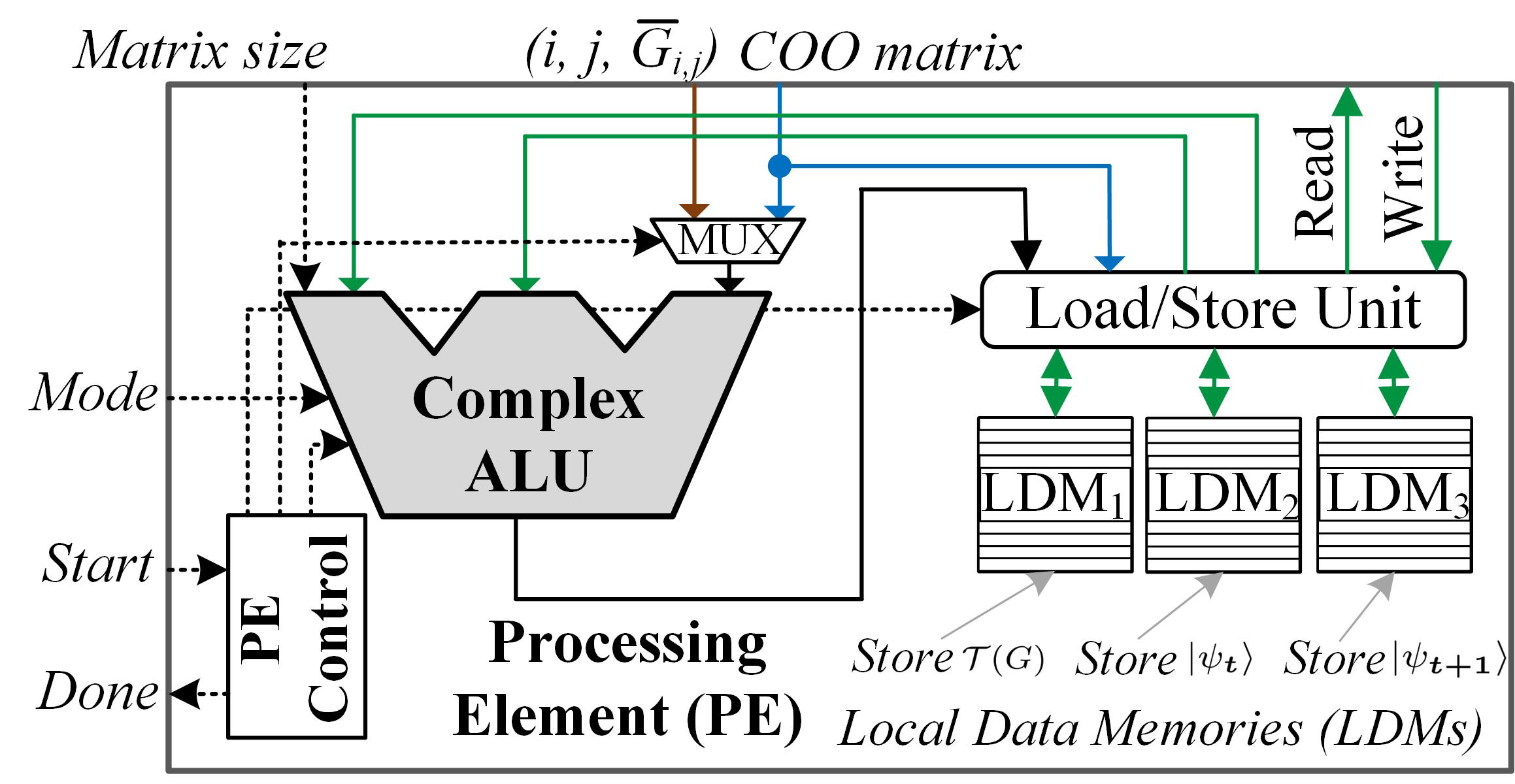}
        \caption{Hardware architecture of the processing element.}
        \label{fig:pe}  
    \end{figure} 

    Fig.~\ref{fig:pe} illustrates the architecture of a Processing Element (PE) designed for matrix operations. The Complex ALU responsible for \textbf{TP} and \textbf{MM} computations will be described in Section~\ref{subsubsec:ALU}. In addition to the basic logic blocks used to synchronize and pipeline the computation between PEs, we introduce three Local Data Memories (LDMs). The computations of $\mathcal{T}(\overline{G})$ and $\mathcal{T}(G)$ are performed sequentially, their results are continuously stored in $\mathcal{T}(\overline{G})$ Memory and $\text{LDM}_{1}$ respectively. After completing the computation of $\mathcal{T}(\overline{G})$, the data in the form of $(i,j,\overline{G}_{i,j})$ is transferred from the $\mathcal{T}(\overline{G})$ Memory to the Complex ALU for the computation of $U$. The state $|\psi^{\bm{t}}\rangle$ is then transferred to the $\text{LDM}_{2}$ to facilitate the computation of the following state, $|\psi^{(\bm{t+1})}\rangle$, which is then stored in $\text{LDM}_{3}$.

    
\subsubsection{Complex ALU for Sparse Matrix Operations}
\label{subsubsec:ALU}

    As outlined in Section~\ref{sec:background}, the efficiency of Quantum Emulation procedures is predominantly dependent on \textbf{TP} and \textbf{MM}, which form the core of all principal operations. Therefore, enhancing the performance of Quantum Emulation is intrinsically linked to the optimization of those computations. 
    
\begin{figure}[t]
        \centering
        \includegraphics[width=0.45\textwidth]{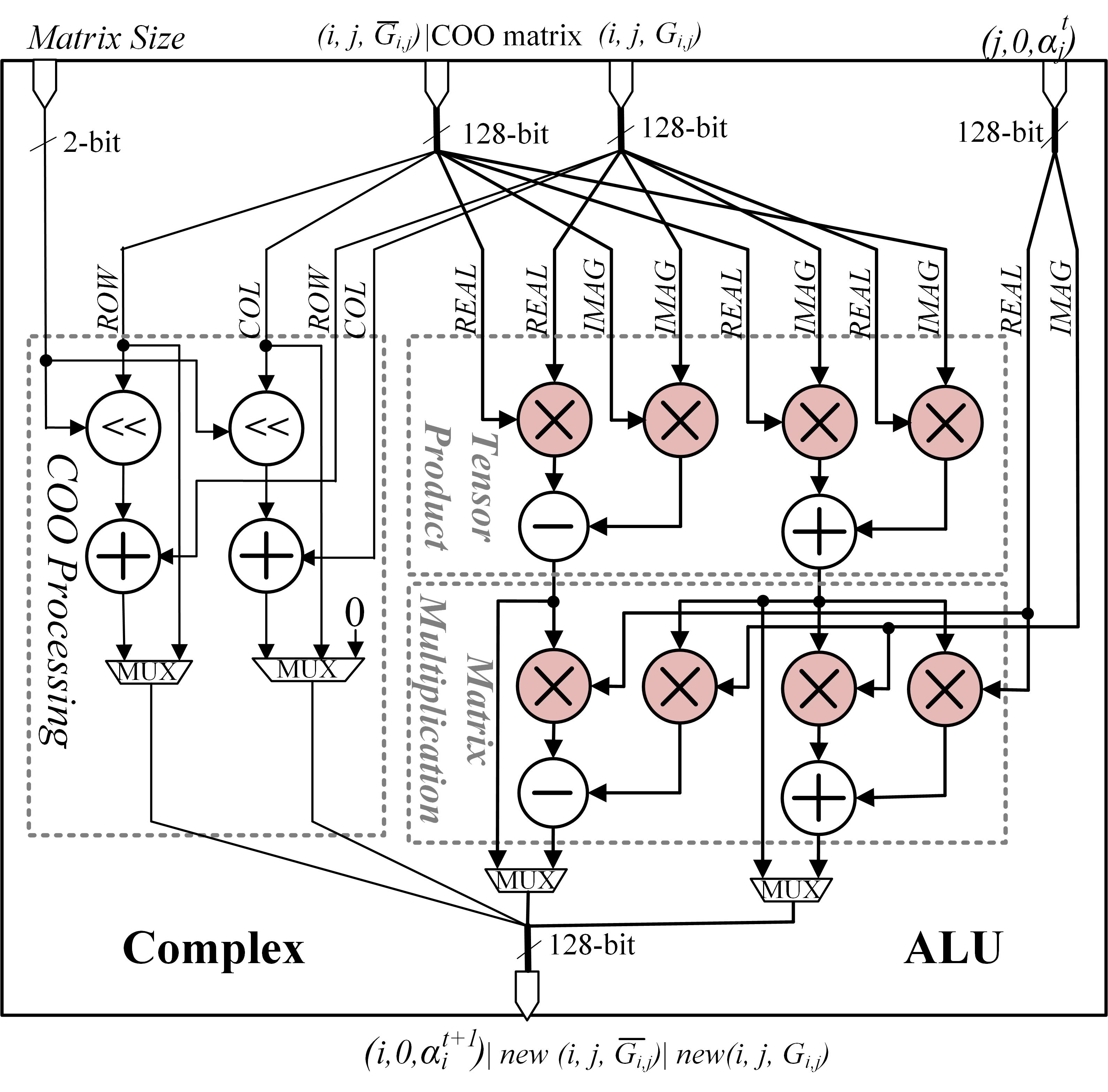}
        \caption{Microarchitecture of the complex ALU.}
        \label{fig:alu}  
    \end{figure}

    Fig.~\ref{fig:alu} illustrates the utilization of the proposed ALU such as COO Processing, \textbf{TP} and \textbf{MM}. Specifically, COO Processing and \textbf{TP} are employed for $\mathcal{T}(\overline{G})$ and $\mathcal{T}(G)$ computations, simultaneously COO Processing, \textbf{TP} and \textbf{MM} utilized to compute $U$ and $|\psi^{(\bm{t})}\rangle$ as described in Section~\ref{subsec:EMMS}. Three 128-bit input tuples $(i,j,\{\text{Re},\text{Im}\})$ belong to $\mathcal{T}(\overline{G})$, $\mathcal{T}(G)$ and $|\psi^{(t)}\rangle$ where $i,j$ are 32-bit unsigned integers, and $\{\text{Re}, \text{Im}\}$ are 32-bit fixed-point real and imaginary parts of a complex number (2 signed integer bits and 30 fractional bits). Additionally, the ``Matrix Size'' signal is the size of the matrix representation of gate $g_j$ ($\hat{N}$), given that our proposed architecture exclusively supports $\hat{N}\in\{2,4\}$. All computational units, including Addition, Multiplication, Subtraction, and Left Shift Operations, are designed in pipelined architecture. This ensures that all matrix computations can be executed with a throughput of approximately one cycle per operation. Four multiplexers are employed to select the output from one of two computational processes.

\section{Analysis and Evaluation}
\label{sec:analysis}

\subsection{Quantitative Analysis of Hardware Resources}
  In this section, we provide theoretical hardware resource estimates for our proposed QEA architecture using the EMMS method, taking into account the maximum ($\#\text{Qubits}$) supported. The analysis focuses on the impact of hardware resources when increasing the number of PEs ($\#\text{PEs}$) and adjusting the depth of the LDM. This approach enables users to select the most suitable architecture for their specific FPGA type. Since the primary costs in QEA development arise from memory usage and multiplication operations, our resource analysis will concentrate on two key components: BRAM and DSP utilization. 
\begin{figure*}[t]
    \centering
    \includegraphics[width=1\textwidth]{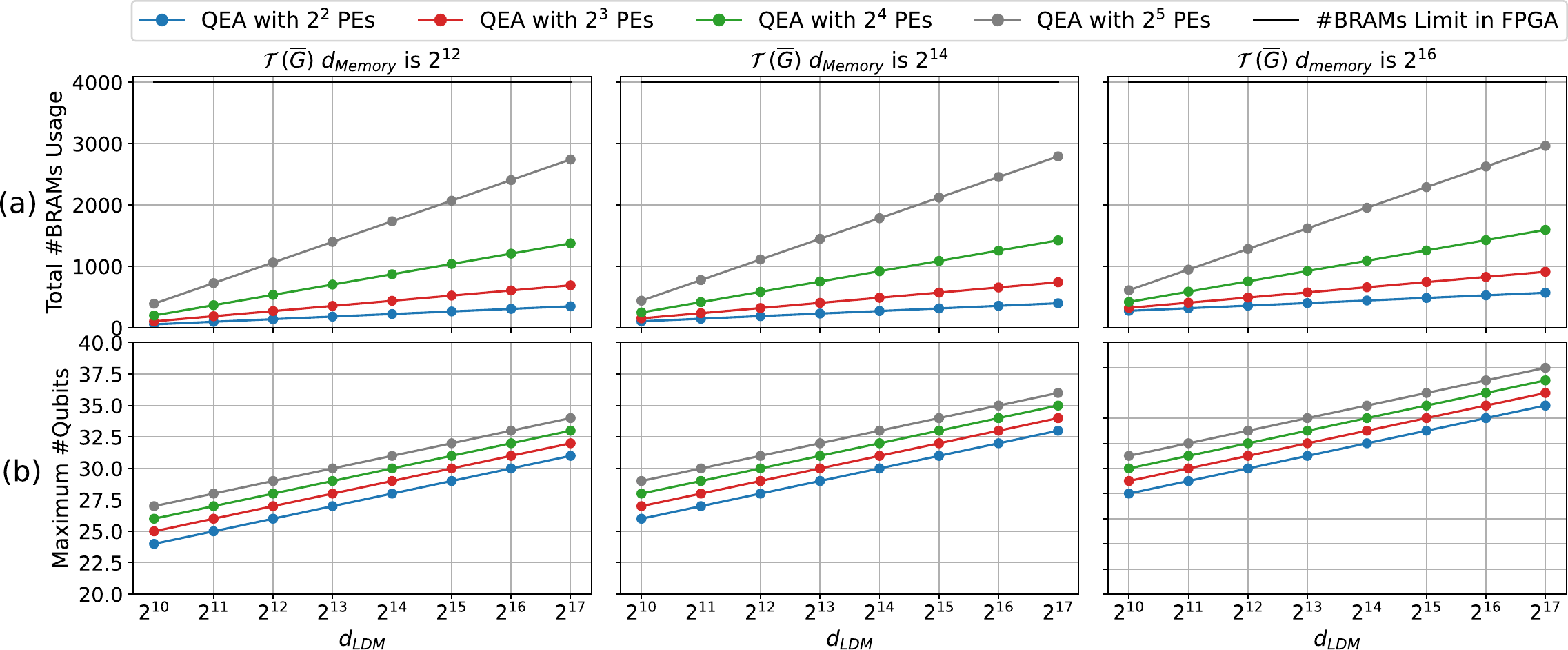}
    \caption{Total estimated $\#$BRAM usage (a) and maximum $\#\text{Qubits}$ support (b) for QEA configurations with $2^2$ to $2^5$ PEs, across $\mathcal{T}(\overline{G})$ Memory depths of $2^{12}$, $2^{14}$, and $2^{16}$.}
    \label{fig:graph_BRAM}  
\end{figure*}

\begin{figure}[t]
    \centering
    \includegraphics[width=0.49\textwidth]{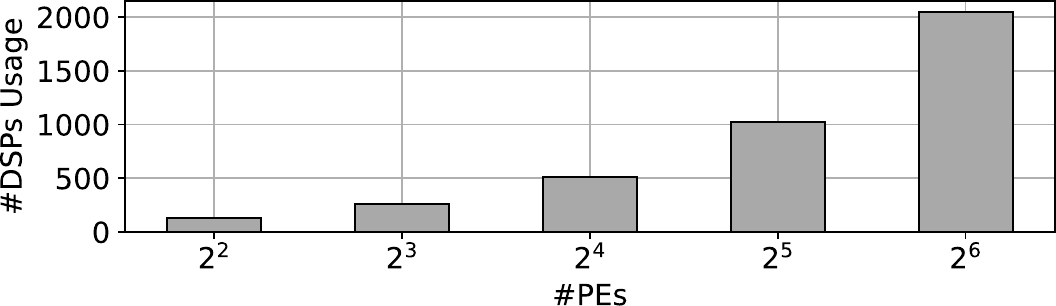}
    \caption{Estimated $\#$DSPs usage for QEA as $\#\text{PEs}$ increases.}
    \label{fig:graph_DSP}  
\end{figure}
  
    Fig.~\ref{fig:graph_BRAM} illustrates the trade-off between BRAM consumption and the maximum $\#\text{Qubits}$ support by QEA configurations with varying $\#\text{PEs}$ and different memory depths of $\mathcal{T}(\overline{G})$. Specifically, as shown in Fig.~\ref{fig:graph_BRAM} (a), BRAM usage increases linearly with the $\#\text{PEs}$ and LDM memory depth, and also grows proportionally with the $\mathcal{T}(\overline{G})$ Memory depth. This relationship demonstrates the scalability of the system, allowing for more PEs at the cost of increased memory to support additional qubits. As shown in Fig.~\ref{fig:graph_BRAM} (b), the maximum $\#\text{Qubits}$ supported increases linearly with the $\#\text{PEs}$, reflecting the doubling of LDM depth as the $\#\text{Qubits}$ grows. This linear trend highlights QEA's scalability in supporting larger quantum circuits with predictable memory growth, enabling the selection of suitable FPGAs easily.
    
   Fig.~\ref{fig:graph_DSP} presents the estimated $\#\text{DSP}$ usage for the QEA architecture as the $\#\text{PEs}$ increases. The eight multipliers of the ALU inside each PE occupy DSP resources. Accordingly, the $\#\text{DSP}$ usage correlates with the increasing $\#\text{PEs}$. At lower PE numbers, such as $2^2$ and $2^3$, the DSP usage remains minimal, under $500$ units. Meanwhile, as the PE number grows to $2^5$ and $2^6$, the DSP usage rises sharply, reaching approximately 2000 units at $2^6$. Even with such high DSP usage, it can still be supported by many FPGAs, indicating the feasibility of increasing the $\#\text{PEs}$ to accelerate the \textbf{TP} and \textbf{MM} calculations. Overall, the EMMS method enables QEA to support up to 32 qubits while maintaining BRAM and DSP resource requirements that can be accommodated by many mid-range FPGAs

\subsection{Quantitative Analysis of Performance}

\begin{figure*}[t]
    \centering
    \includegraphics[width=1\textwidth]{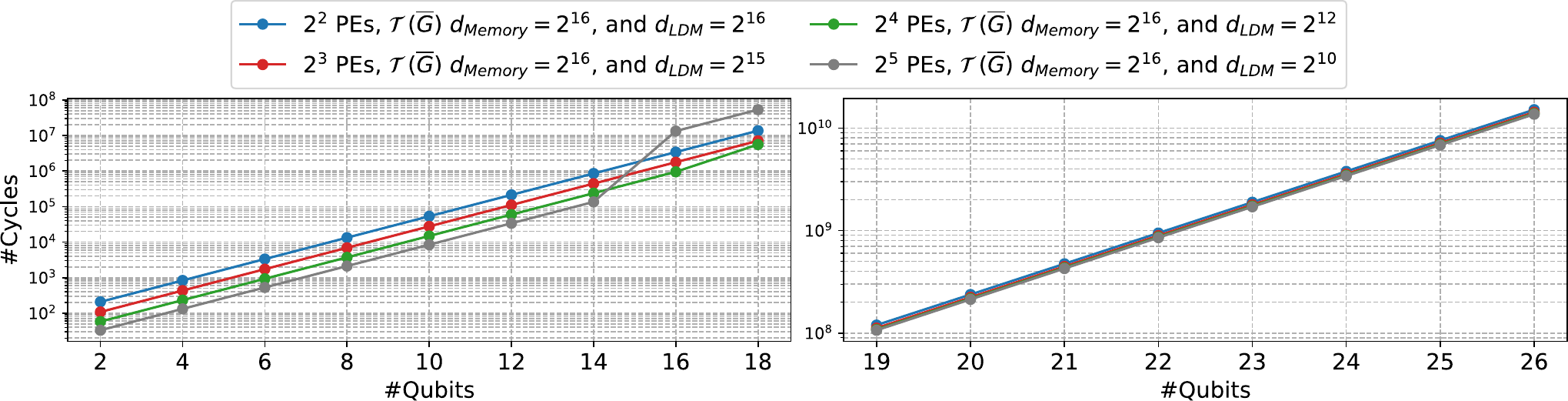}
    \caption{Estimated cycle count of four different versions of QEA on the Xilinx ZCU102 FPGA, required for performing $100$ ($m=100$) computations for the target quantum state across various $\#\text{Qubits}$.}
    \label{fig:graph_cycle}  
\end{figure*}

This section analyzes the system-level performance of QEA to clarify how the EMMS method and the $\#\text{PEs}$ affect processing efficiency. To accurately estimate performance, we evaluate QEA on the Xilinx Zynq UltraScale+ ZCU102 FPGA. Based on the available resources of the ZCU102 FPGA, we selected the four best versions of QEA, with $2^2$ to $2^5$ PEs, each configured with the same $\mathcal{T}(\overline{G})$ Memory depth of $2^16$ and different LDM depths. Since this paper focuses on theoretical analysis and no actual hardware has been designed using Verilog, we assess performance based on the number of execution cycles.

For benchmarking, we choose a quantum computation model with a quantum circuit that involves \textit{M} computations for the target quantum state $|\psi^{(\bm{m})}\rangle$. In cases where the $\#\text{Qubits}$ used are fewer than the total LDM depth of all PEs, the number of execution cycles for the quantum circuit ($\text{Cycle}_{\text{QC}}$), which can compute state vectors consecutively in $\text{LDM}_2$ and $\text{LDM}_3$ without continuous transmission, can be calculated as Eq.~ \ref{eq:cqc_1}:

\begin{equation}
\text{C}_{\text{QC}} = \text{C}_{\text{Write } |\psi^{(\bm{0})}\rangle} + \sum_{t=0}^{m-1} \left( \text{C}^{U^{(\bm{t+1})}}_{\textbf{TP}} + \text{C}^{U^{(\bm{t+1})}}_{\textbf{MM}} \right) + \text{C}_{\text{Read } |\psi^{(\bm{m})}\rangle}
\label{eq:cqc_1}
\end{equation}

Where $\text{C}_{\text{Write} |\psi^{(\bm{0})}\rangle}$ is the number of cycles ($\#\text{Cycles}$) of the first $|\psi^{(\bm{0})}\rangle$ transmission, $\text{C}^{U^{(\bm{t+1})}}_{\textbf{TP}}$ is $\#\text{Cycles}$ for computing the \textbf{TP} for $U^{(\bm{t+1})}$, $\text{C}^{U^{(\bm{t+1})}}_{\textbf{MM}}$ is the $\#\text{Cycles}$ for computing the matrix multiplication of $U^{(\bm{t+1})}$ with $|\psi^{(\bm{t})}\rangle$, and $\text{C}_{\text{Read } |\psi^{(\bm{m})}\rangle}$ is $\#\text{Cycles}$ for reading final $|\psi^{(\bm{m})}\rangle$ result. According to our experimental results on the ZCU102 FPGA, the high-performance port from the processing system to QEA allows 128-bit transmission and reception per cycle, which is equivalent to one 128-bit COO value of an element $\alpha^{(\bm{j})}$ in $|\psi^{(\bm{t})}\rangle$. As a result, $\text{C}_{\text{Write } |\psi^{(\bm{t})}\rangle}$ and $\text{C}_{\text{Read } |\psi^{(\bm{t+1})}\rangle}$ are equal to $N$, where $n$ is the used $\#\text{Qubits}$. Meanwhile, $\text{C}^{U^{(\bm{t+1})}}_{\textbf{TP}}$ and $\text{C}^{U^{(\bm{t+1})}}_{\textbf{MM}}$ can be shortened by paralleling the computation through increasing the $\#\text{PEs}$, as calculated in Eq.~ \ref{eq:ctp_cmm}:

\begin{equation}
\text{C}^{U^{(\bm{t+1})}}_{\textbf{TP}} = (\overline{N} + N/\overline{N})/\#\text{PEs}; \text{C}^{U^{(\bm{t+1})}}_{\textbf{MM}} = N/\#\text{PEs}.
\label{eq:ctp_cmm}
\end{equation}

In cases where the $\#\text{Qubits}$ used are larger than the total LDM depth of all PEs, $\text{C}_{\text{QC}}$ increases dramatically because the computed state vectors $|\psi^{(\bm{t})}\rangle$ and $|\psi^{(\bm{t+1})}\rangle$ are only partially stored in $\text{LDM}_2$ and $\text{LDM}_3$. This requires the system to continuously write a portion of $|\psi^{(\bm{t})}\rangle$ and read a portion of $|\psi^{(\bm{t+1})}\rangle$ from DDRAM to the QEA, and vice versa, to complete the update of the entire $|\psi^{(\bm{t+1})}\rangle$. As a result, $\text{C}_{\text{Write } |\psi^{(\bm{t})}\rangle}$ and $\text{C}_{\text{Read } |\psi^{(\bm{t+1})}\rangle}$ become obvious bottlenecks in the system, causing $\text{C}_{\text{QC}}$ to increase rapidly following the \textit{M} computations, as calculated in Eq.~\ref{eq:cqc_2}.

\begin{equation}
\resizebox{0.49\textwidth}{!}{$
\text{C}_{\text{QC}} = \sum_{\bm{t}=0}^{m-1} \left( \text{C}_{\text{Write } |\psi^{(\bm{t})}\rangle} + \text{C}^{U^{(\bm{t+1})}}_{\textbf{TP}} + \text{C}^{U^{(\bm{t+1})}}_{\textbf{MM}} + \text{C}_{\text{Read } |\psi^{(\bm{t+1})}\rangle}\right) 
\label{eq:cqc_2}
$}\end{equation}

Based on the analysis in Eq.~\ref{eq:cqc_1} and \ref{eq:cqc_2}, we estimate the number of computation cycles for the four QEA versions for $100$ ($m=100$) computations of the target quantum state with $\#\text{Qubits}$ ranging from 2 to 26, as shown in Fig.~\ref{fig:graph_cycle}. Note that the Xilinx ZCU102 FPGA has only 4GB of DDR4 memory, so the maximum number of effectively supported $\#\text{Qubits}$ is limited to 26, as storing $|\psi^{(\bm{t})}\rangle$ and $|\psi^{(\bm{t+1})}\rangle$ for 26 qubits requires 2GB. Accordingly, from 2 to 18 qubits, $|\psi^{(\bm{t})}\rangle$ and $|\psi^{(\bm{t+1})}\rangle$ can be stored entirely in the $\text{LDM}_2$ and $\text{LDM}_3$ of the PEs, so the $\#\text{Cycles}$ for the QEA versions decreases as the $\#\text{PEs}$ increases. The exceptions are the QEA version with $2^5$ PEs and LDM depths of $2^{10}$, which performs best for fewer than 15 qubits, and the QEA version with $2^4$ PEs and LDM depths of $2^{12}$, which performs best for fewer than 16 qubits. Beyond 19 qubits, increasing the $\#\text{PEs}$ does not significantly reduce the total number of computation cycles, as $\text{C}_{\text{Write } |\psi^{(\bm{t})}\rangle}$ and $\text{C}_{\text{Read } |\psi^{(\bm{t+1})}\rangle}$ account for 66\% of the processing time. As a result, the $\#\text{Cycles}$ for all four QEA versions becomes largely insignificant and depends primarily on the transmission bandwidth between DDR4 and the QEA.

Overall, the EMMS method makes QEA computationally feasible for large $\#\text{Qubits}$. For computations with fewer than 18 qubits, increasing the $\#\text{PEs}$ significantly accelerates performance. However, for more than 18 qubits, system performance becomes heavily dependent on the transmission bandwidth between the QEA and DDRAM. This limitation can be addressed by adopting modern FPGAs with high-bandwidth memory support, which would improve data transfer rates and overall system performance.

\subsection{Comparison with State-of-the-Art Works}

\begin{figure}[t]
    \centering
    \includegraphics[width=0.5\textwidth]{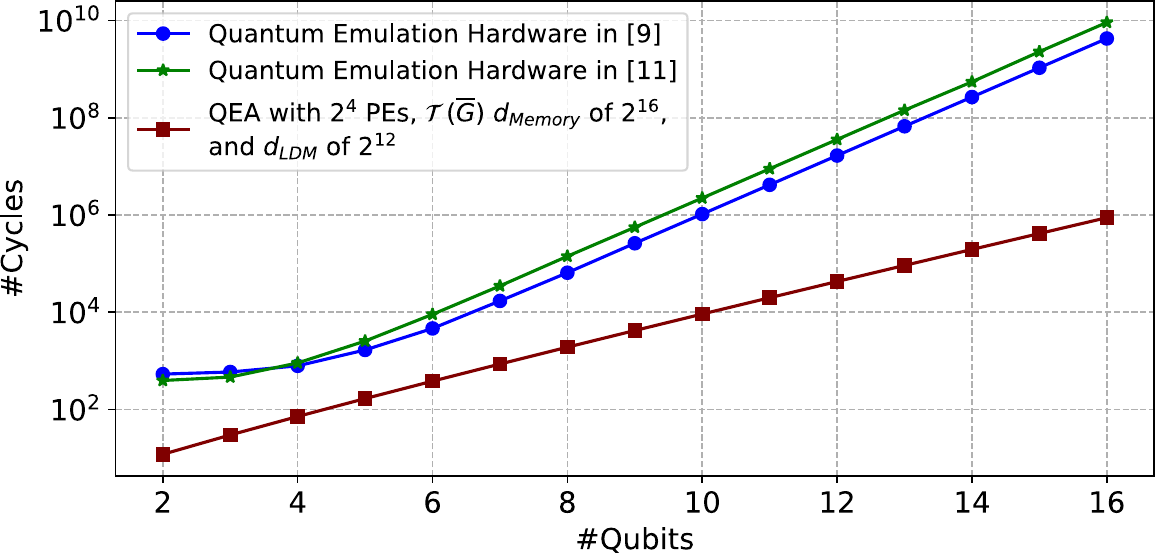}
    \caption{$\#\text{Cycles}$ between QEA and other FPGA-based quantum emulation from \cite{mahmud2020efficient,Mahmud2020Grover} on QFT algorithm.}
    \label{fig:graph_comparison}  
    \vspace{-5mm}
\end{figure} 

This section compares the performance of QEA with quantum emulation hardware studies. To our knowledge, no quantum emulation hardware with the gate fusion approach is available to date, so the performance of the EMMS method cannot be compared with FPGA-based works. Instead, we compare the performance of QEA with quantum emulation hardware dedicated to updating traditional quantum state vectors, as described in \cite{mahmud2020efficient,Mahmud2020Grover}, which does not use gate fusion. For a fair comparison, we convert the execution time reported in \cite{mahmud2020efficient,Mahmud2020Grover} into cycles and benchmark the QFT application.

Fig.~\ref{fig:graph_comparison} compares the computational cycles of three hardware architectures across qubits ranging from 2 to 16 in QFT applications. The two hardware designs in \cite{mahmud2020efficient,Mahmud2020Grover} exhibit a steep increase in cycle count as the $\#\text{Qubits}$ grows. This is because these designs do not utilize the COO storage method, leading to exponential growth in computational complexity as the size of $U^{(\bm{t})}$ expands with the $\#\text{Qubits}$. In contrast, by virtue of the COO storage method, our QEA implementation with 16 PEs shows a significant reduction in the $\#\text{Cycles}$ across all qubit counts compared to \cite{mahmud2020efficient,Mahmud2020Grover}. Besides, the QEA effectively leverages parallelism by increasing the $\#\text{PEs}$ to 16, resulting in a 16$\times$ acceleration in computations. Overall, this comparison highlights the dramatic impact of COO storage and PE-based parallelism on reducing computational cycles.

\section{Conclusion \& Future Work} \label{sec:conclusion}

This study introduced the EMMS method and QEA architecture to address memory and computational challenges in quantum emulation. By leveraging multiple PEs, the QEA significantly accelerates tensor product and matrix multiplication tasks. Theoretical analysis on the Xilinx ZCU102 FPGA shows linear scalability with qubit count and substantial reductions in computation cycles for circuits with fewer than 18 qubits, outperforming previous designs. However, for circuits exceeding 19 qubits, performance becomes bottlenecked by the data transfer bandwidth between DDR4 memory and the QEA, which consumes 66\% of the total execution time. 

In future work, since QEA has only been evaluated theoretically, we will design QEA in Verilog and implement it practically on the Xilinx ZCU102 FPGA. Additionally, we will implement QEA on the Alveo U280 FPGA, which features high-bandwidth memory of up to 460 GB/s, to address the bottleneck issue in data transmission and reception for calculations involving more than 18 qubits.



\section*{Acknowledgment}

This work was supported by JST-ALCA-Next Program Grant Number JPMJAN23F5, Japan. The research has been partly executed in response to the support of JSPS, KAKENHI Grant No. 22H00515, Japan.

\bibliographystyle{IEEEtran}
\bibliography{references.bib}

\section*{Appendix A: Example for Eq.~\ref{eq:explicit}}
\label{sec:example}


Consider $X$-gate act on $0^{\text{th}}$ qubit in $3$-qubits system $|\psi^{(\bm{t})}\rangle$. We know that $\mathcal{T}(\overline{G})=X_0=\{
(0,1,1),
(1,0,1)\}$ and $\mathcal{T}(G)=\mathcal{T}(I_1\otimes I_2)=\{
(0,0,1),
(1,1,1),
(2,2,1),
(3,3,1)
\}$. Following Algo.~\ref{algo:basic}, with $\overline{N}=2$, at $i = 0$ and $i = 1$:

\begin{equation}
    \begin{split}
        &\left[\overline{G}_{0, 1}\times \mathcal{T}(G)\right] \times |\psi^{(\bm{t})}\rangle_{4:7} = |\psi^{(\bm{t+1})}\rangle_{0:3},\\
        &\left[\overline{G}_{1, 0}\times \mathcal{T}(G)\right] \times |\psi^{(\bm{t})}\rangle_{0:3} = |\psi^{(\bm{t+1})}\rangle_{4:7},
    \end{split}
\end{equation}

respectively, the final result obtained through concatenation.

\section*{Appendix B: Compressed gates}

The detailed list of quantum gates in COO format has been shown in Table~\ref{tab:c_gate}.

\begin{table}[]
\renewcommand{\arraystretch}{1.2}
\caption{Compressed sparse gates (first $11$ rows) and dense gates (last $2$ rows) in COO format.}
\label{tab:c_gate}
\resizebox{0.49\textwidth}{!}{
\begin{tabular}{cccccccccc}
\hline
\multicolumn{1}{|c|}{\multirow{2}{*}{\textbf{Gate}}}          & \multicolumn{4}{c|}{\textbf{COO}}                                                                                                              & \multicolumn{1}{c|}{\multirow{2}{*}{\textbf{Gate}}}          & \multicolumn{4}{c|}{\textbf{COO}}                                                                                                              \\ \cline{2-5} \cline{7-10} 
\multicolumn{1}{|c|}{}                                        & \multicolumn{1}{c|}{\textbf{$i$}} & \multicolumn{1}{c|}{\textbf{$j$}} & \multicolumn{1}{c|}{\textbf{Re}} & \multicolumn{1}{c|}{\textbf{Im}} & \multicolumn{1}{c|}{}                                        & \multicolumn{1}{c|}{\textbf{$i$}} & \multicolumn{1}{c|}{\textbf{$j$}} & \multicolumn{1}{c|}{\textbf{Re}} & \multicolumn{1}{c|}{\textbf{Im}} \\ \hline
\multicolumn{1}{|c|}{\multirow{2}{*}{\textit{P($\lambda$)}}}  & \multicolumn{1}{c|}{0}            & \multicolumn{1}{c|}{1}            & \multicolumn{1}{c|}{1}             & \multicolumn{1}{c|}{0}            & \multicolumn{1}{c|}{\multirow{4}{*}{\textit{CX}}}            & \multicolumn{1}{c|}{0}            & \multicolumn{1}{c|}{0}            & \multicolumn{1}{c|}{1}             & \multicolumn{1}{c|}{0}            \\ \cline{2-5} \cline{7-10} 
\multicolumn{1}{|c|}{}                                        & \multicolumn{1}{c|}{1}            & \multicolumn{1}{c|}{1}            & \multicolumn{1}{c|}{$a^{*}$}       & \multicolumn{1}{c|}{$b^{*}$}      & \multicolumn{1}{c|}{}                                        & \multicolumn{1}{c|}{1}            & \multicolumn{1}{c|}{1}            & \multicolumn{1}{c|}{1}             & \multicolumn{1}{c|}{0}            \\ \cline{1-5} \cline{7-10} 
\multicolumn{1}{|c|}{\multirow{2}{*}{\textit{X}}}             & \multicolumn{1}{c|}{0}            & \multicolumn{1}{c|}{1}            & \multicolumn{1}{c|}{1}             & \multicolumn{1}{c|}{0}            & \multicolumn{1}{c|}{}                                        & \multicolumn{1}{c|}{2}            & \multicolumn{1}{c|}{3}            & \multicolumn{1}{c|}{1}             & \multicolumn{1}{c|}{0}            \\ \cline{2-5} \cline{7-10} 
\multicolumn{1}{|c|}{}                                        & \multicolumn{1}{c|}{1}            & \multicolumn{1}{c|}{0}            & \multicolumn{1}{c|}{1}             & \multicolumn{1}{c|}{0}            & \multicolumn{1}{c|}{}                                        & \multicolumn{1}{c|}{3}            & \multicolumn{1}{c|}{2}            & \multicolumn{1}{c|}{1}             & \multicolumn{1}{c|}{0}            \\ \hline
\multicolumn{1}{|c|}{\multirow{2}{*}{\textit{Y}}}             & \multicolumn{1}{c|}{0}            & \multicolumn{1}{c|}{1}            & \multicolumn{1}{c|}{0}             & \multicolumn{1}{c|}{$-1$}           & \multicolumn{1}{c|}{\multirow{4}{*}{\textit{CY}}}            & \multicolumn{1}{c|}{0}            & \multicolumn{1}{c|}{0}            & \multicolumn{1}{c|}{1}             & \multicolumn{1}{c|}{0}            \\ \cline{2-5} \cline{7-10} 
\multicolumn{1}{|c|}{}                                        & \multicolumn{1}{c|}{1}            & \multicolumn{1}{c|}{0}            & \multicolumn{1}{c|}{0}             & \multicolumn{1}{c|}{1}            & \multicolumn{1}{c|}{}                                        & \multicolumn{1}{c|}{1}            & \multicolumn{1}{c|}{1}            & \multicolumn{1}{c|}{1}             & \multicolumn{1}{c|}{0}            \\ \cline{1-5} \cline{7-10} 
\multicolumn{1}{|c|}{\multirow{2}{*}{\textit{Z}}}             & \multicolumn{1}{c|}{0}            & \multicolumn{1}{c|}{0}            & \multicolumn{1}{c|}{1}             & \multicolumn{1}{c|}{0}            & \multicolumn{1}{c|}{}                                        & \multicolumn{1}{c|}{2}            & \multicolumn{1}{c|}{3}            & \multicolumn{1}{c|}{1}             & \multicolumn{1}{c|}{$-1$}           \\ \cline{2-5} \cline{7-10} 
\multicolumn{1}{|c|}{}                                        & \multicolumn{1}{c|}{1}            & \multicolumn{1}{c|}{1}            & \multicolumn{1}{c|}{$-1$}            & \multicolumn{1}{c|}{0}            & \multicolumn{1}{c|}{}                                        & \multicolumn{1}{c|}{3}            & \multicolumn{1}{c|}{2}            & \multicolumn{1}{c|}{1}             & \multicolumn{1}{c|}{1}            \\ \hline
\multicolumn{1}{|c|}{\multirow{2}{*}{\textit{S}}}             & \multicolumn{1}{c|}{0}            & \multicolumn{1}{c|}{0}            & \multicolumn{1}{c|}{1}             & \multicolumn{1}{c|}{0}            & \multicolumn{1}{c|}{\multirow{4}{*}{\textit{CZ}}}            & \multicolumn{1}{c|}{0}            & \multicolumn{1}{c|}{0}            & \multicolumn{1}{c|}{1}             & \multicolumn{1}{c|}{0}            \\ \cline{2-5} \cline{7-10} 
\multicolumn{1}{|c|}{}                                        & \multicolumn{1}{c|}{1}            & \multicolumn{1}{c|}{1}            & \multicolumn{1}{c|}{0}             & \multicolumn{1}{c|}{1}            & \multicolumn{1}{c|}{}                                        & \multicolumn{1}{c|}{1}            & \multicolumn{1}{c|}{1}            & \multicolumn{1}{c|}{1}             & \multicolumn{1}{c|}{0}            \\ \cline{1-5} \cline{7-10} 
\multicolumn{1}{|c|}{\multirow{2}{*}{\textit{SDG}}}           & \multicolumn{1}{c|}{0}            & \multicolumn{1}{c|}{0}            & \multicolumn{1}{c|}{1}             & \multicolumn{1}{c|}{0}            & \multicolumn{1}{c|}{}                                        & \multicolumn{1}{c|}{2}            & \multicolumn{1}{c|}{2}            & \multicolumn{1}{c|}{1}             & \multicolumn{1}{c|}{0}            \\ \cline{2-5} \cline{7-10} 
\multicolumn{1}{|c|}{}                                        & \multicolumn{1}{c|}{1}            & \multicolumn{1}{c|}{1}            & \multicolumn{1}{c|}{0}             & \multicolumn{1}{c|}{$-1$}           & \multicolumn{1}{c|}{}                                        & \multicolumn{1}{c|}{3}            & \multicolumn{1}{c|}{3}            & \multicolumn{1}{c|}{$-1$}            & \multicolumn{1}{c|}{0}            \\ \hline
\multicolumn{1}{|c|}{\multirow{2}{*}{\textit{T}}}             & \multicolumn{1}{c|}{0}            & \multicolumn{1}{c|}{0}            & \multicolumn{1}{c|}{1}             & \multicolumn{1}{c|}{0}            & \multicolumn{1}{c|}{\multirow{4}{*}{\textit{CP($\lambda$)}}} & \multicolumn{1}{c|}{0}            & \multicolumn{1}{c|}{0}            & \multicolumn{1}{c|}{1}             & \multicolumn{1}{c|}{0}            \\ \cline{2-5} \cline{7-10} 
\multicolumn{1}{|c|}{}                                        & \multicolumn{1}{c|}{1}            & \multicolumn{1}{c|}{1}            & \multicolumn{1}{c|}{$q^{*}$}             & \multicolumn{1}{c|}{$q^{*}$}            & \multicolumn{1}{c|}{}                                        & \multicolumn{1}{c|}{1}            & \multicolumn{1}{c|}{1}            & \multicolumn{1}{c|}{1}             & \multicolumn{1}{c|}{0}            \\ \cline{1-5} \cline{7-10} 
\multicolumn{1}{|c|}{\multirow{2}{*}{\textit{TDG}}}           & \multicolumn{1}{c|}{0}            & \multicolumn{1}{c|}{0}            & \multicolumn{1}{c|}{1}             & \multicolumn{1}{c|}{0}            & \multicolumn{1}{c|}{}                                        & \multicolumn{1}{c|}{2}            & \multicolumn{1}{c|}{2}            & \multicolumn{1}{c|}{1}             & \multicolumn{1}{c|}{0}            \\ \cline{2-5} \cline{7-10} 
\multicolumn{1}{|c|}{}                                        & \multicolumn{1}{c|}{1}            & \multicolumn{1}{c|}{1}            & \multicolumn{1}{c|}{$-q^{*}$}            & \multicolumn{1}{c|}{$q^{*}$}            & \multicolumn{1}{c|}{}                                        & \multicolumn{1}{c|}{3}            & \multicolumn{1}{c|}{3}            & \multicolumn{1}{c|}{$a^{*}$}       & \multicolumn{1}{c|}{$b^{*}$}      \\ \hline
\multicolumn{1}{|c|}{\multirow{2}{*}{\textit{RZ($\theta$)}}}  & \multicolumn{1}{c|}{0}            & \multicolumn{1}{c|}{0}            & \multicolumn{1}{c|}{$c^{*}$}            & \multicolumn{1}{c|}{$-d^{*}$}          & \multicolumn{1}{c|}{\multirow{6}{*}{\textit{CRY($\theta$)}}} & \multicolumn{1}{c|}{0}            & \multicolumn{1}{c|}{0}            & \multicolumn{1}{c|}{1}             & \multicolumn{1}{c|}{0}            \\ \cline{2-5} \cline{7-10} 
\multicolumn{1}{|c|}{}                                        & \multicolumn{1}{c|}{1}            & \multicolumn{1}{c|}{1}            & \multicolumn{1}{c|}{$c^{*}$}            & \multicolumn{1}{c|}{$d^{*}$}           & \multicolumn{1}{c|}{}                                        & \multicolumn{1}{c|}{1}            & \multicolumn{1}{c|}{1}            & \multicolumn{1}{c|}{1}             & \multicolumn{1}{c|}{0}            \\ \cline{1-5} \cline{7-10} 
\multicolumn{1}{|c|}{\multirow{4}{*}{\textit{CRZ($\theta$)}}} & \multicolumn{1}{c|}{0}            & \multicolumn{1}{c|}{0}            & \multicolumn{1}{c|}{1}             & \multicolumn{1}{c|}{0}            & \multicolumn{1}{c|}{}                                        & \multicolumn{1}{c|}{2}            & \multicolumn{1}{c|}{2}            & \multicolumn{1}{c|}{$c^{*}$}            & \multicolumn{1}{c|}{0}            \\ \cline{2-5} \cline{7-10} 
\multicolumn{1}{|c|}{}                                        & \multicolumn{1}{c|}{1}            & \multicolumn{1}{c|}{1}            & \multicolumn{1}{c|}{1}             & \multicolumn{1}{c|}{0}            & \multicolumn{1}{c|}{}                                        & \multicolumn{1}{c|}{2}            & \multicolumn{1}{c|}{3}            & \multicolumn{1}{c|}{$-d^{*}$}           & \multicolumn{1}{c|}{0}            \\ \cline{2-5} \cline{7-10} 
\multicolumn{1}{|c|}{}                                        & \multicolumn{1}{c|}{2}            & \multicolumn{1}{c|}{2}            & \multicolumn{1}{c|}{$c^{*}$}            & \multicolumn{1}{c|}{$-d^{*}$}          & \multicolumn{1}{c|}{}                                        & \multicolumn{1}{c|}{3}            & \multicolumn{1}{c|}{2}            & \multicolumn{1}{c|}{$d^{*}$}            & \multicolumn{1}{c|}{0}            \\ \cline{2-5} \cline{7-10} 
\multicolumn{1}{|c|}{}                                        & \multicolumn{1}{c|}{3}            & \multicolumn{1}{c|}{3}            & \multicolumn{1}{c|}{$c^{*}$}            & \multicolumn{1}{c|}{$d^{*}$}           & \multicolumn{1}{c|}{}                                        & \multicolumn{1}{c|}{3}            & \multicolumn{1}{c|}{3}            & \multicolumn{1}{c|}{$c^{*}$}            & \multicolumn{1}{c|}{0}            \\ \hline
\multicolumn{1}{|c|}{\multirow{6}{*}{\textit{CRX($\theta$)}}} & \multicolumn{1}{c|}{0}            & \multicolumn{1}{c|}{0}            & \multicolumn{1}{c|}{1}             & \multicolumn{1}{c|}{0}            & \multicolumn{1}{c|}{\multirow{6}{*}{\textit{CH}}}            & \multicolumn{1}{c|}{0}            & \multicolumn{1}{c|}{0}            & \multicolumn{1}{c|}{1}             & \multicolumn{1}{c|}{0}            \\ \cline{2-5} \cline{7-10} 
\multicolumn{1}{|c|}{}                                        & \multicolumn{1}{c|}{1}            & \multicolumn{1}{c|}{1}            & \multicolumn{1}{c|}{1}             & \multicolumn{1}{c|}{0}            & \multicolumn{1}{c|}{}                                        & \multicolumn{1}{c|}{1}            & \multicolumn{1}{c|}{1}            & \multicolumn{1}{c|}{1}             & \multicolumn{1}{c|}{0}            \\ \cline{2-5} \cline{7-10} 
\multicolumn{1}{|c|}{}                                        & \multicolumn{1}{c|}{2}            & \multicolumn{1}{c|}{\textit{2}}   & \multicolumn{1}{c|}{$c^{*}$}            & \multicolumn{1}{c|}{0}            & \multicolumn{1}{c|}{}                                        & \multicolumn{1}{c|}{2}            & \multicolumn{1}{c|}{2}            & \multicolumn{1}{c|}{$q^{*}$}             & \multicolumn{1}{c|}{0}            \\ \cline{2-5} \cline{7-10} 
\multicolumn{1}{|c|}{}                                        & \multicolumn{1}{c|}{2}            & \multicolumn{1}{c|}{3}            & \multicolumn{1}{c|}{0}             & \multicolumn{1}{c|}{$-d^{*}$}          & \multicolumn{1}{c|}{}                                        & \multicolumn{1}{c|}{2}            & \multicolumn{1}{c|}{3}            & \multicolumn{1}{c|}{$q^{*}$}             & \multicolumn{1}{c|}{0}            \\ \cline{2-5} \cline{7-10} 
\multicolumn{1}{|c|}{}                                        & \multicolumn{1}{c|}{3}            & \multicolumn{1}{c|}{2}            & \multicolumn{1}{c|}{0}             & \multicolumn{1}{c|}{$-d^{*}$}          & \multicolumn{1}{c|}{}                                        & \multicolumn{1}{c|}{3}            & \multicolumn{1}{c|}{2}            & \multicolumn{1}{c|}{$q^{*}$}             & \multicolumn{1}{c|}{0}            \\ \cline{2-5} \cline{7-10} 
\multicolumn{1}{|c|}{}                                        & \multicolumn{1}{c|}{3}            & \multicolumn{1}{c|}{3}            & \multicolumn{1}{c|}{$c^{*}$}            & \multicolumn{1}{c|}{0}            & \multicolumn{1}{c|}{}                                        & \multicolumn{1}{c|}{3}            & \multicolumn{1}{c|}{3}            & \multicolumn{1}{c|}{$-q^{*}$}            & \multicolumn{1}{c|}{0}            \\ \hline\hline
\multicolumn{1}{|c|}{\multirow{4}{*}{\textit{H}}}             & \multicolumn{1}{c|}{0}            & \multicolumn{1}{c|}{0}            & \multicolumn{1}{c|}{$q^{*}$}       & \multicolumn{1}{c|}{0}            & \multicolumn{1}{c|}{\multirow{4}{*}{\textit{RX($\theta$)}}}  & \multicolumn{1}{c|}{0}            & \multicolumn{1}{c|}{0}            & \multicolumn{1}{c|}{$c^{*}$}       & \multicolumn{1}{c|}{0}            \\ \cline{2-5} \cline{7-10} 
\multicolumn{1}{|c|}{}                                        & \multicolumn{1}{c|}{0}            & \multicolumn{1}{c|}{1}            & \multicolumn{1}{c|}{$q^{*}$}       & \multicolumn{1}{c|}{0}            & \multicolumn{1}{c|}{}                                        & \multicolumn{1}{c|}{0}            & \multicolumn{1}{c|}{1}            & \multicolumn{1}{c|}{0}             & \multicolumn{1}{c|}{$-d^{*}$}     \\ \cline{2-5} \cline{7-10} 
\multicolumn{1}{|c|}{}                                        & \multicolumn{1}{c|}{1}            & \multicolumn{1}{c|}{0}            & \multicolumn{1}{c|}{$q^{*}$}       & \multicolumn{1}{c|}{0}            & \multicolumn{1}{c|}{}                                        & \multicolumn{1}{c|}{1}            & \multicolumn{1}{c|}{0}            & \multicolumn{1}{c|}{0}             & \multicolumn{1}{c|}{$-d^{*}$}     \\ \cline{2-5} \cline{7-10} 
\multicolumn{1}{|c|}{}                                        & \multicolumn{1}{c|}{1}            & \multicolumn{1}{c|}{1}            & \multicolumn{1}{c|}{$-q^{*}$}      & \multicolumn{1}{c|}{0}            & \multicolumn{1}{c|}{}                                        & \multicolumn{1}{c|}{1}            & \multicolumn{1}{c|}{1}            & \multicolumn{1}{c|}{$q^{*}$}       & \multicolumn{1}{c|}{0}            \\ \hline
\multicolumn{1}{|c|}{\multirow{4}{*}{\textit{SX}}}            & \multicolumn{1}{c|}{0}            & \multicolumn{1}{c|}{0}            & \multicolumn{1}{c|}{$p^{*}$}       & \multicolumn{1}{c|}{$p^{*}$}      & \multicolumn{1}{c|}{\multirow{4}{*}{\textit{RY($\theta$)}}}  & \multicolumn{1}{c|}{0}            & \multicolumn{1}{c|}{0}            & \multicolumn{1}{c|}{$c^{*}$}       & \multicolumn{1}{c|}{0}            \\ \cline{2-5} \cline{7-10} 
\multicolumn{1}{|c|}{}                                        & \multicolumn{1}{c|}{0}            & \multicolumn{1}{c|}{1}            & \multicolumn{1}{c|}{$p^{*}$}       & \multicolumn{1}{c|}{$-p^{*}$}     & \multicolumn{1}{c|}{}                                        & \multicolumn{1}{c|}{0}            & \multicolumn{1}{c|}{1}            & \multicolumn{1}{c|}{$-d^{*}$}      & \multicolumn{1}{c|}{0}            \\ \cline{2-5} \cline{7-10} 
\multicolumn{1}{|c|}{}                                        & \multicolumn{1}{c|}{1}            & \multicolumn{1}{c|}{0}            & \multicolumn{1}{c|}{$p^{*}$}       & \multicolumn{1}{c|}{$-p^{*}$}     & \multicolumn{1}{c|}{}                                        & \multicolumn{1}{c|}{1}            & \multicolumn{1}{c|}{0}            & \multicolumn{1}{c|}{$d^{*}$}       & \multicolumn{1}{c|}{0}            \\ \cline{2-5} \cline{7-10} 
\multicolumn{1}{|c|}{}                                        & \multicolumn{1}{c|}{1}            & \multicolumn{1}{c|}{1}            & \multicolumn{1}{c|}{$p^{*}$}       & \multicolumn{1}{c|}{$p^{*}$}      & \multicolumn{1}{c|}{}                                        & \multicolumn{1}{c|}{1}            & \multicolumn{1}{c|}{1}            & \multicolumn{1}{c|}{$c^{*}$}       & \multicolumn{1}{c|}{0}            \\ \hline
\multicolumn{10}{l}{$a:\cos(\lambda), b^*:\sin(\lambda),c^{*}:\cos(\theta/2), d^{*}:\sin(\theta/2), q^{*}: 1/\sqrt{2}, p^{*}: 1/2$}                                                                                                        
\end{tabular}}
\end{table}

\end{document}